\documentstyle[12pt]{article}

\begin{document}
\begin{titlepage}
\begin{center}

March 16, 2000    \hfill    LBNL-45229 \\

\vskip .5in

{\large \bf Decoherence, Quantum Zeno Effect, and the Efficacy of Mental
Effort.}
\footnote{This work is supported in part by the Director, Office of Science, 
Office of High Energy and Nuclear Physics, Division of High Energy Physics, 
of the U.S. Department of Energy under Contract DE-AC03-76SF00098}

\vskip .20in
Henry P. Stapp\\
{\em Lawrence Berkeley National Laboratory\\
      University of California\\
    Berkeley, California 94720}
\end{center}

\vskip .10in
\begin{abstract}
Recent theoretical and experimental papers support the prevailing 
opinion that large warm systems will rapidly lose quantum coherence, 
and that classical properties will emerge. This rapid loss of coherence 
would naturally be expected to block any critical role for quantum theory 
in explaining the interaction between our conscious experiences and  the 
physical activities of our brains. However, there is a quantum theory of 
mind in which the efficacy of mental effort is not affected by decoherence 
effects. In this theory the effects of mental action on brain activity is 
achieved by a Quantum Zeno Effect that is not weakened by decoherence. The 
theory is based on a relativistic version of von Neumann's quantum theory. 
It encompasses all the predictions of Copenhagen quantum theory, which include
all the validated predictions of classical physical theory. In addition, it 
forges two-way dynamical links between the physical and experiential aspects 
of nature. The theory has significant explanatory power.

\end{abstract}
\medskip
\end{titlepage}

\newpage
\renewcommand{\thepage}{\arabic{page}}
\setcounter{page}{1}

\noindent {\bf 1. Introduction.}

The experimental work of the Paris group of S. Haroche [1] and of the 
Boulder group of D. Wineland [2] demonstrate convincingly that the theoretical
ideas of quantum theory really do work in careful experiments
performed, in effect, on individual atoms interacting with  
controlled electromagnetic probes and environments. It is
an impressive tribute to the power of human reason and logic that the
creators of quantum theory were able to accurately forecast
effects so far removed in scale and intricacy from the data
that they possessed. 

The experiments of these groups both confirm the 
emergence of decoherence effects whose strength and rapidity of onset
increase rapidly with the size of the system being disturbed
by interactions with its environment.

In recent theoretical paper [3] Max Tegmark computes, on the basis of the 
thus-confirmed ideas, some expected time intervals for the disappearance of 
quantum coherence in various brain structures that have been proposed  as 
the seat of the neural correlates of consciousness. He finds that quantum
coherence disappears on time scales of $10^{-13}$ to $10^{-20}$ seconds, 
and concludes from this that classical concepts should provide a
completely adequate basis for understanding the dynamical
connection between mind and brain. 

This conclusion depends on the idea that the quantum interaction between 
mind and brain depends upon quantum coherence. It is indeed usually thought 
that coherence is the essence of quantum theory, and that all quantum effects 
depend upon it. But the development of the von Neumann-Wigner quantum theory 
of mind pursued by this author was specifically designed so that the effect of
mental effort on brain process is not weakened by decoherence. Indeed, 
quantum decoherence was {\it assumed} to decompose the state of the brain 
into a mixture of essentially classical states. But the quantum effect of 
mental effort on brain activity is not curtailed by this decomposition. 

I shall now explain how this works. 

{\bf 2. Overview of the Theory}

Before giving the specific computation I must first describe 
the general form of the theory. It is based on objectively interpreted 
von Neumann-Wigner quantum theory. I have argued elsewhere [4,5] that 
the evolving state S(t) of von Neumann-Wigner quantum theory can be 
construed to be our theoretical representation of an objectively existing
and evolving informational structure that can properly be called ``physical 
reality''.  

The theory has four basic equations. The first defines the 
state of a subsystem. If S(t) is the operator that represents 
the state of the universe and b is a subsystem of the 
universe then the state of b is defined to be
$$
S(t)_b = Tr_b S(t),    \eqno(2.1)         
$$
where $Tr_b$ means the trace over all variable except 
those that characterize b.

The second basic equation specifies von Neumann's process I. 
This process ``poses a question''. If
$S(t-0)$ represents the limit of $S(t')$ as $t'$ approaches t from 
below then at certain times t the following jump occurs:
$$
 S(t)= P S(t-0) P + (1-P) S(t-0) (1-P). \eqno(2.2) 
$$
Here P is a projection operator (i.e., $P^2 = P$) that acts as 
the unit operator on all degrees of freedom except those 
associated with the processor b.

The third basic equation specifies the (Dirac) reduction. This
reduction specifies nature's answer to the question:
$$
S(t+0)= P S(t) P \mbox{ with probability } Tr P S(t)/ Tr S(t)  \eqno (2.3)\\
$$
or
$$
S(t+0)=(1-P) S(t) (1-P) \mbox{ with probability }Tr (1-P) S(t)/Tr S(t).
$$

Between jumps the state evolves according to:
$$
S(t+\Delta t)= \exp(-iH\Delta t) S(t) \exp(+iH\Delta t). \eqno (2.4)
$$

The projection operator P has two eigenvalues, 1 and 0, and is
therefore  associated
with a Yes-No question: the two alternative possible 
reductions specified 
in (2.3) are associated with the two alternative possible 
answers, Yes or No, to the question associated with P. Thus 
the reduction (2.3) specifies 
one bit of information, and implants that information in the 
state S(t) of the physical universe. This state S(t) can be 
regarded as just the evolving carrier of the bits of 
information generated by these reduction events.

Information is normally conceived to be associated with an 
interpreting system. In Copenhagen quantum theory each 
reduction is associated with an increment in human knowledge, 
and the interpreting system is the brain and body of the 
observer. Generalizing from this one known
kind of example, I shall assume that each reduction (2.3) is 
associated with a quantum information processor, call it b, 
that both poses the question
---picks P---and, when nature responds by picking, say, the 
answer P=1, `interprets' that bit of information by 
evolving in a characteristic way. 

The projection operator P cannot be local: any point-like 
projection would inject infinite energy into the processor.
This jump of S(t) to P S(t)P, because it is basically a 
nonlocal process, has no counterpart in classical dynamics: 
it is a new kind of element, relative to classical physical 
theory. Generalizing again from the one known example, I 
assume that each reduction event is connected to some sort 
of ``knowing'': each such event has a characteristic 
experiential ``feel''.

Each thought involves an effort to attend to something--- 
i.e., to pose a question---followed by a registration of the 
answer. This conforms exactly to the quantum dynamics.

Normally a sequence of thoughts consists of a string of 
thoughts each of which differs just slightly from its 
predecessor: the sequence becomes a `stream' of 
consciousness. So the basic process is self-replication:
the thought T creates conditions that tend to create a 
likeness of T.
 
This means that a key requirement for P is that PSP not
evolve rapidly out of the subspace defined by P, or at least
that PSP quickly evolve into a state nearly the same as PSP,
so that the sequence of thought is likely to be a sequence of
similar thoughts.

One possibility is that the projection operator P may act in 
the space of a set of conjugate variables that is undergoing 
periodic motion, and that it projects onto a band of neighboring 
orbits in phase space. For a simple harmonic oscillator in 
a state of high energy one could take the projection operator 
P to be the sum of the projection operators onto a large set 
of neighboring energy eigenstates. This would effectively 
project onto a band of neighboring orbits in phase space.

\noindent {\bf 3. The Quantum Zeno Effect}

In this theory the main effect of mind on brain is via the 
quantum Zeno effect. Suppose the initial state is PS(t)P, 
and that in that state the next question is again P, and that 
this question repetitiously repeats.
If these questions are posed at intervals $\Delta t$ then
equations (2.4) and (2.2) give
$$
S(t+\Delta t) = P \exp (-iH\Delta t) PS(t)P \exp (+iH\Delta t) P  
$$
$$
+ (1-P) \exp (-iH\Delta t) PS(t)P \exp (+iH\Delta t) (1-P).
$$
If $\Delta t$ is small on the scale of the leakage of PS(t)P
out of the subspace defined by P then the second term is
small and of second order in $\Delta t$. Thus as $\Delta t$
gets small, on the scale of the leakage of $PSP$ into the subspace
associated with $(1-P)$, the Hamiltonian $H$ gets effectively 
replaced by $PHP$: evolution within the $P$ subspace proceeds 
normally, but leakage out of that subspace is blocked.

The point here is that the linear-in-time leakage out of the 
subspace defined by $P$ is killed by the reduction events. 
Thus only the quadratic and higher terms survive, and these 
are damped out if the reductions occurs fast on the time 
scale of the relevant oscillations.

This replacement of the full Hamiltonian H by PHP is the usual
quantum Zeno effect. We see that it is just as effective 
for a statistical mixture S(t) of quasi-classical states as 
for a pure state: the decoherence generated by interaction
with the environment does not weaken this quantum effect.

\noindent {\bf 4. Explanatory Power}

Von Neumann-Wigner quantum theory 
encompasses all the valid predictions of classical physical 
theory. So for any computation, or argumentation, for which 
quantum effects are unimportant one can use classical physics.
Hence vN/W theory is at least as good as classical physical 
theory: the two theories are effectively equivalent insofar as
quantum effects are unimportant. In the purely physical 
domain the vN/W theory is certainly better, because it 
predicts also all of the quantum effects, including all of 
the ``nonlocal'' quantum effects. But our interest here 
is on the nature of the dynamical link between mind and brain,
and the nature of the consequences of this connection.

The only power given to the mind by this theory is the power  
to choose the questions P. And the only effects of these 
choices that has thus far been identified are the consequences
achieved by the quantum Zeno effect. This effect is to keep
the brain activity focussed on a question for longer than 
it would stay focussed in the classical theory.

To make the theory still more constrained, let me assume that
the quantum processor, in  this case the human brain/body, 
possesses a certain set of possible questions P, and that 
at a prescribed  sequence of instants the processor can either 
consent, or not consent, to posing a certain possible question P.
Let this question P be the one that maximizes $Tr_b P S(T)/Tr_b S(t)$. 
To accomodate our intuitive feeling that mental `effort' does effect 
brain/body activity I add the postulate that the rapidity of 
the sequence of instants can be increased by mental effort.  

This is a simple theory. But the effect of mind on brain
is highly constrained. The only variables under mental control
are ``consent' and `effort'.

Does this theory explain anything?

Consider the following passage from ``Psychology: 
The Briefer Course'' by William James [7]. In the final 
section of the chapter on Attention he
writes:

``I have spoken as if our attention were wholly 
determined by neural conditions. I believe that the array of {\it things}
we can attend to is so determined. No object can {\it catch} our attention
except by the neural machinery. But the {\it amount} of the attention which
an object receives after it has caught our attention is another question.
It often takes effort to keep mind upon it. We feel that we can make more 
or less of the effort as we choose. If this feeling be not deceptive, 
if our effort be a spiritual force, and an indeterminant one, then of 
course it contributes coequally with the cerebral conditions to the result.
Though it introduce no new idea, it will deepen and prolong the stay in 
consciousness of innumerable ideas which else would fade more quickly
away. The delay thus gained might not be more than a second in duration---
but that second may be critical; for in the rising and falling 
considerations in the mind, where two associated systems of them are
nearly in equilibrium it is often a matter of but a second more or 
less of attention at the outset, whether one system shall gain force to
occupy the field and develop itself and exclude the other, or be excluded 
itself by the other. When developed it may make us act, and that act may 
seal our doom. When we come to the chapter on the Will we shall see that 
the whole drama of the voluntary life hinges on the attention, slightly 
more or slightly less, which rival motor ideas may receive. ...''  

Posing a question is the act of attending. In the chapter on Will, in the
section entitled ``Volitional effort is effort of  attention'' [7]
James writes:

``Thus  we find that {\it we reach the  heart  of our inquiry  into volition
when we ask by what process is it that the thought of any given action
comes to prevail stably in the mind.}'' 

and later

``{\it  The essential achievement of the will, in short, when it is most 
`voluntary,'  is to attend to a difficult  object and hold it fast before
the  mind.   ...  Effort of attention is  thus the essential phenomenon
of will.''}

Still  later, James says:

{\it  ``Consent to the idea's undivided presence, this is effort's sole 
achievement.''} ...``Everywhere, then, the function  of effort is the same:
to keep affirming and adopting the thought  which,  if left to  itself, would 
slip away.''
  
The vN/W theory, with the quantum zeno effect incorporated, 
explains naturally the features that are the basis of James's 
conception of the action of human volition.

\noindent {\bf References}

1. M. Brune, et. al. Phys. Rev. Lett. {\bf 77}, 4887 (1996)

2. C.J. Myatt, et. al. Nature, {\bf 403}, 269 (2000)
 
3. Max Tegmark, ``The Importance of Quantum Decoherence in Brain
   Process,''  Phys. Rev E, to appear.

4. H.P. Stapp, ``Nonlocality, Counterfactuals, and Consistent Histories,\\
   http://xxx.lanl.gov/abs/quant-ph/9905055

5. H.P. Stapp, ``From Einstein Nonlocality to Von Neumann Reality,''\\
   http://www-physics.lbl.gov/$\sim$stapp/stappfiles.html\\
   quant-ph/0003064 

6. H.P. Stapp, ``Attention, Intention, and Will in Quantum Physics,''\\
   in J. Consc. Studies {\bf 6}, 143-64 (1999).

7. Wm. James, ``Psychology: The  Briefer Course'', ed. Gordon Allport,
  University of Notre Dame Press, Notre Dame, IN.  Ch. 4 and Ch. 17

\end{document}